\documentclass[letterpaper, 10 pt, conference]{ieeeconf}                                                             
\IEEEoverridecommandlockouts                            
\usepackage{graphics} 
\usepackage{amsmath}
\usepackage{amssymb} 
\usepackage{graphicx}
\usepackage{subcaption}
\usepackage{color,colortbl}
\usepackage{url}
\newtheorem{thm}{Assumption}

\title{\textbf{Improvement in the UAV position estimation with low-cost GPS, INS and vision-based system: Application to a quadrotor UAV}}

\author{L.Arreola, A. Montes de Oca, A. Flores, J. Sanchez, G. Flores
\thanks{Corresponding author: G. Flores (email: gflores@cio.mx). L. Arreola, A. Montes de Oca, A. Flores, J. Santis and G. Flores are with the Perception and Robotics Laboratory, Optics Research Center, Le\'{o}n 37150, Guanajuato, Mexico, 37150.}
\thanks{This work was supported by the FORDECYT-CONACYT under agreement 000000000292399.}
}

\begin{document}
\maketitle
\thispagestyle{empty}
\pagestyle{empty}
\begin{abstract}
In this paper, we develop a position estimation system for Unmanned Aerial Vehicles formed by hardware and software. It is based on low-cost devices: GPS, commercial autopilot sensors and dense optical flow algorithm implemented in an onboard microcomputer. Comparative tests were conducted using our approach and the conventional one, where only fusion of GPS and inertial sensors are used. Experiments were conducted using a quadrotor in two flying modes: hovering and trajectory tracking in outdoor environments. Results demonstrate the effectiveness of the proposed approach in comparison with the conventional approaches presented in the vast majority of commercial drones.
\end{abstract}
\IEEEpeerreviewmaketitle
\section{Introduction}
In many outdoor Unmanned Aerial Vehicles (UAV aka drones) applications such as: inspection \& monitoring, mapping, precision agriculture and civil engineering, just to mention a few, precision in the drone position estimation is crucial. Hence commercial drones use to use high-precision Global Positioning System (GPS) or even Real-Time Kinematic (RTK) devices \cite{7500933}, \cite{8115945}. In this paper an easy-to-implement application is proposed as a low-cost option to the commercial technology available in the market. Such implementation is based on the combination of two popular technologies: conventional low-cost GPS and a camera-based system endowed with optical flow algorithm. On one hand, GPS provides to drone the capability to follow a specific trajectory in global coordinates, however low-cost GPS have intrinsic sources that induce errors in the UAV position. This results in an insufficient accuracy for several of the aforementioned applications. On the other hand, optical flow devices provide information about UAV relative position w.r.t. a fixed reference frame. However, optical flow devices by itself are unable to provide global coordinates, which are necessary in the vast majority of outdoor applications, for instance in 3D reconstruction and mapping. 
%
\begin{figure}[t]\vspace{0.35cm}
  \centering
    \includegraphics[width=0.45\textwidth]{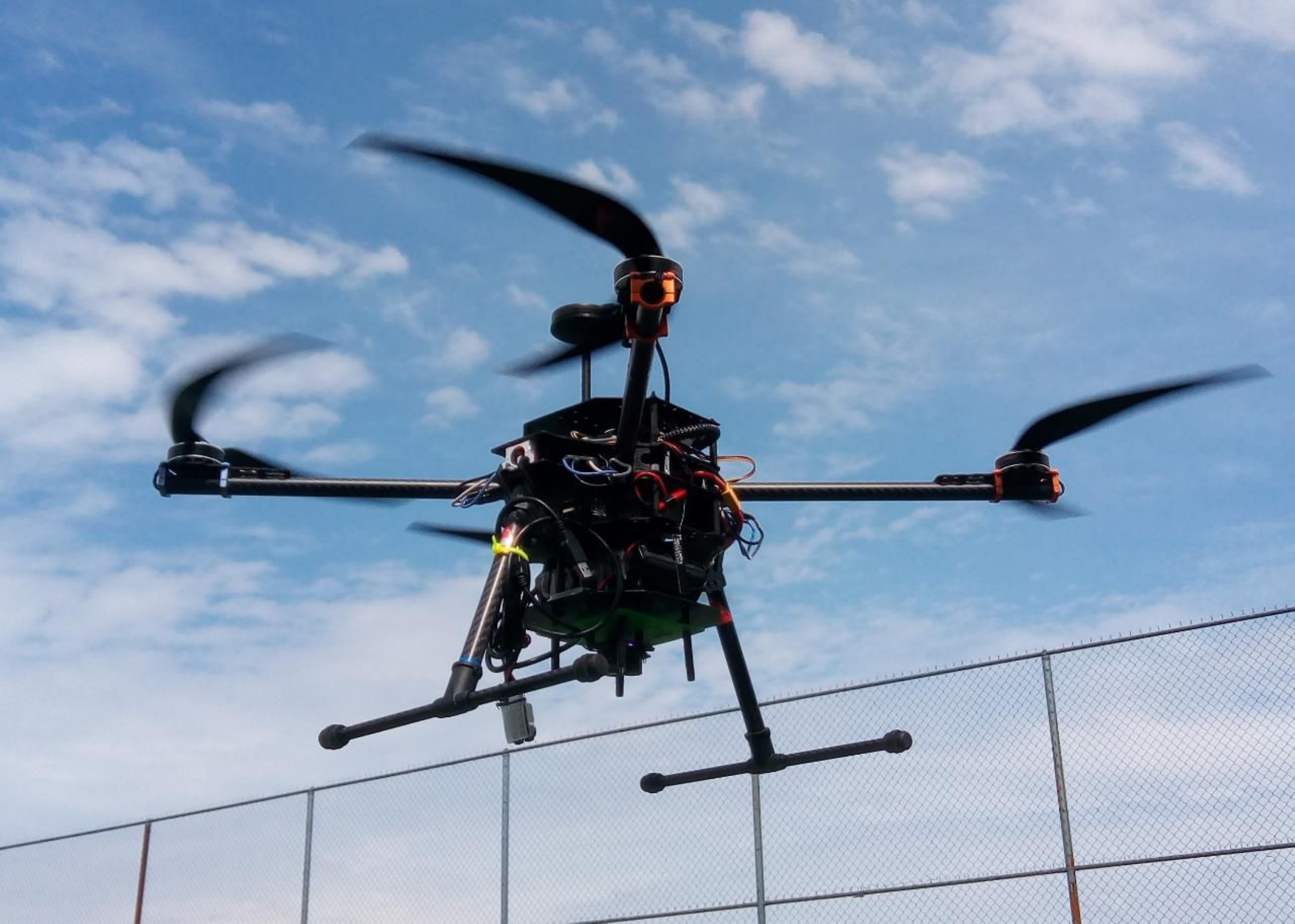}
  \caption{The quadrotor unmanned aerial vehicle used in this study. It has a camera at the bottom connected to a microcomputer to compute optical flow algorithm on-board.}
  \label{drone}
\end{figure}

The vast majority of the works available in the literature shows recent developments in the position estimation in denied GPS environments \cite{4543471}, \cite{5152680}, \cite{7738795}, \cite{WCICA14:Hong}. A comparative study of optical flow and a combination of measurements from GPS-INS sensors through UAV flight tests is presented in \cite{ACC13:Chao}, where inaccuracies in the optical flow are evident compared with the position provided by the GPS. The group of the authors of the last paper continues their research in \cite{IROS14:Rhudy}, where they present an Unscented Information Filter for estimating the vehicle speed of GPS-denied navigation. In that work the authors use GPS measurements as reference values for the fusion algorithm using INS and optical flow information. In \cite{Fusini:ICUAS15} a nonlinear observer for UAV velocity based on INS and optical flow data is implemented. The results do not show visible improvements in the accuracy of UAV position estimation w.r.t classical filter approaches. One of the most recent works presented at \cite{Qi:ICUS17}, proposes a fusion of the PX4Flow sensor \cite{ICRA13:Flow} and INS system; the authors show some results with considerable error in position estimation (around $5$ meters) from the desired trajectory, which is insufficient for various applications. The use of RTK or PPK (Post-Processing Kinematic) systems for global navigation satellite systems (GNSS) nowadays are the most accurate for outdoor applications, this is due to the accuracy they own, which ranges between $20$mm and $50$mm \cite{tractor}, \cite{ppkvsrtk}. Although these technologies are one of the best options in the market for UAV position estimation in outdoor environments, they can be very expensive due to the amount of equipment needed compared to low-cost commercial drones.

In this paper we have conducted several outdoor experiments in the task of trajectory tracking and hovering at a given outdoor environment. Such experiments are compared with the classical approach where only GPS and inertial navigation sensors (INS) information are used to estimate the current UAV position. Such experiments show that the accuracy of the proposed system improves considerably UAV position estimation, and hence the accuracy in tracking a given trajectory. This works continues a research presented in \cite{Mercado:ICUAS13}, where a GPS/INS/Optic flow data fusion for position and velocity estimation is presented. In such a work the algorithm presents a considerable error in the estimation, also it has not been compared with any well-known approach. The contribution of this work w.r.t to our past paper is: a) the implementation of our approach in a quadrotor shown at Fig. \ref{drone}; b) improvement in the optical flow algorithm; and c) enhancement of the data fusion algorithm. The presented results are compared with the data obtained by the classical Kalman Filter, which uses only GPS and INS information. This is provided by the PX4 firmware for quadrotors, a firmware very popular nowadays. Also, experiments were conducted in hovering and in trajectory tracking, in both cases the results are highly satisfactory in comparison with classical approaches presented in the literature. The results we have obtained demonstrate a considerable efficiency w.r.t. to standard GPS and INS systems; such approaches usually have position errors around $2.5$m \cite{stgps}, which results in an insufficient accuracy for certain applications. Therefore our approach results in a viable alternative for limited budget.

The remainder of this paper is organized as follows. In Section \ref{sec:problem} it is described the general approach and the architecture of the system. Section \ref{sec:exp} presents flight experiments with a quadrotor, in hover and in trajectory tracking, by using the proposed approach to estimate the quadrotor position. Finally, Section \ref{sec:conclusion} presents some concluding remarks and an outline of future directions of the presented research.
%
\section{System description} 
\label{sec:problem}
In this section we describe the main software components of the quadrotor system. Also the hardware description of such a system is explained next.
%
\subsection{Hardware}
The quadrotor used in this paper is shown at Fig. \ref{drone}. This drone is equipped with a flight controller, a GPS, a camera rig, four ESCs, a set of four motors and their corresponding propellers, a Li-ion battery, telemetry kit, a microcomputer Odroid XU4 and a voltage regulator. The quadrotor specifications are listed in Table \ref{table:UAVdescription}.
\begin{table}[ht]
\centering
\begin{tabular}{llr} 
\hline
\multicolumn{2}{c}{UAV design} \\
\cline{1-2}
Parameter & Value\\
\hline
Span      & 70 [cm]  \\
Height    & 26 [cm]       \\
Weight    & $\approx$ 2 [kg] \\
Propulsion& Brushless motor 330 [kv]    \\
        & Propeller 17x5.5 [in]\\
Max. Load & $\approx$ 3.5 [kg]\\
Battery type   & Li-Ion 6s\\
                 & Capacity 9500 [mAh]    \\
Flight controller & Pixhack v3 \\
Firmware  & ArduCopter 3.5.5 \\
Estimated flight time & 15 [min]\\
\hline
\end{tabular}
    \caption{Quadrotor UAV parameters.}
    \label{table:UAVdescription}
\end{table}
%
%
\subsection{Dense optical flow algorithm}
\begin{figure}[htb!]
\centering
	\begin{subfigure}{0.23\textwidth}
    	\includegraphics[width=0.95\textwidth]{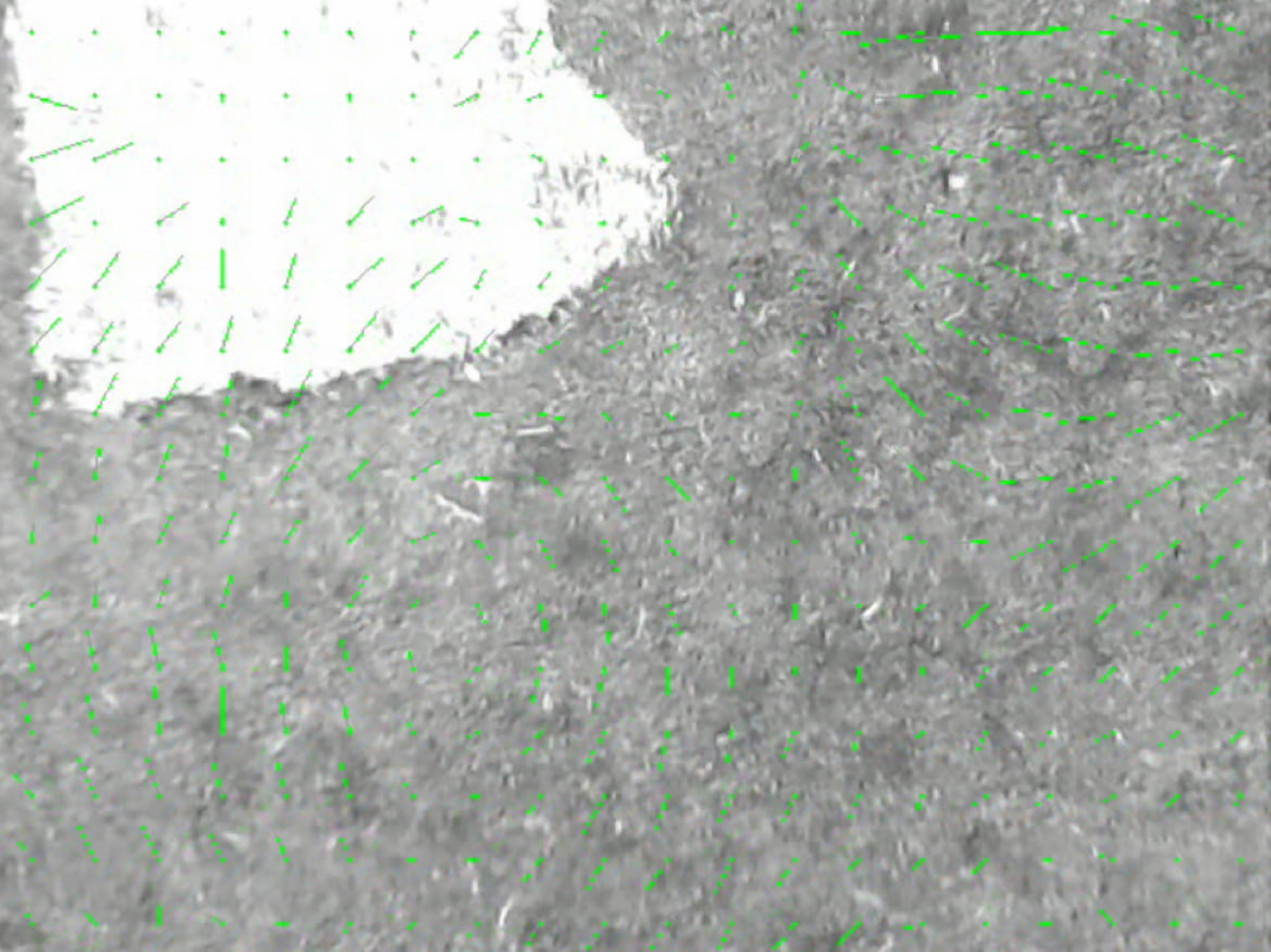}
        \caption{}
	\end{subfigure}
\hfill
	\begin{subfigure}{0.23\textwidth}
    	\includegraphics[width=0.95\textwidth]{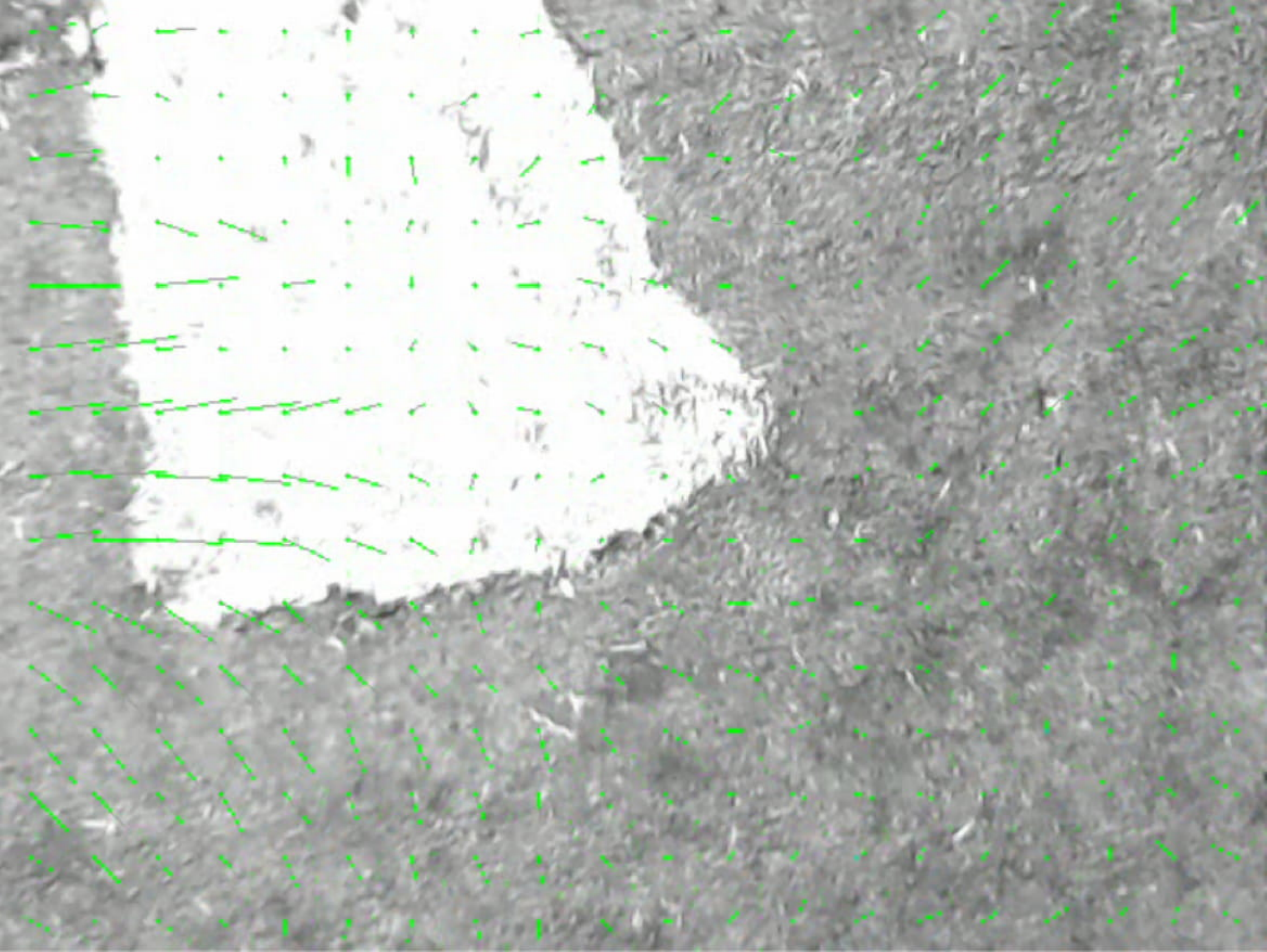}
        \caption{}
	\end{subfigure}
\vfill
	\begin{subfigure}{0.23\textwidth}
    	\includegraphics[width=0.95\textwidth]{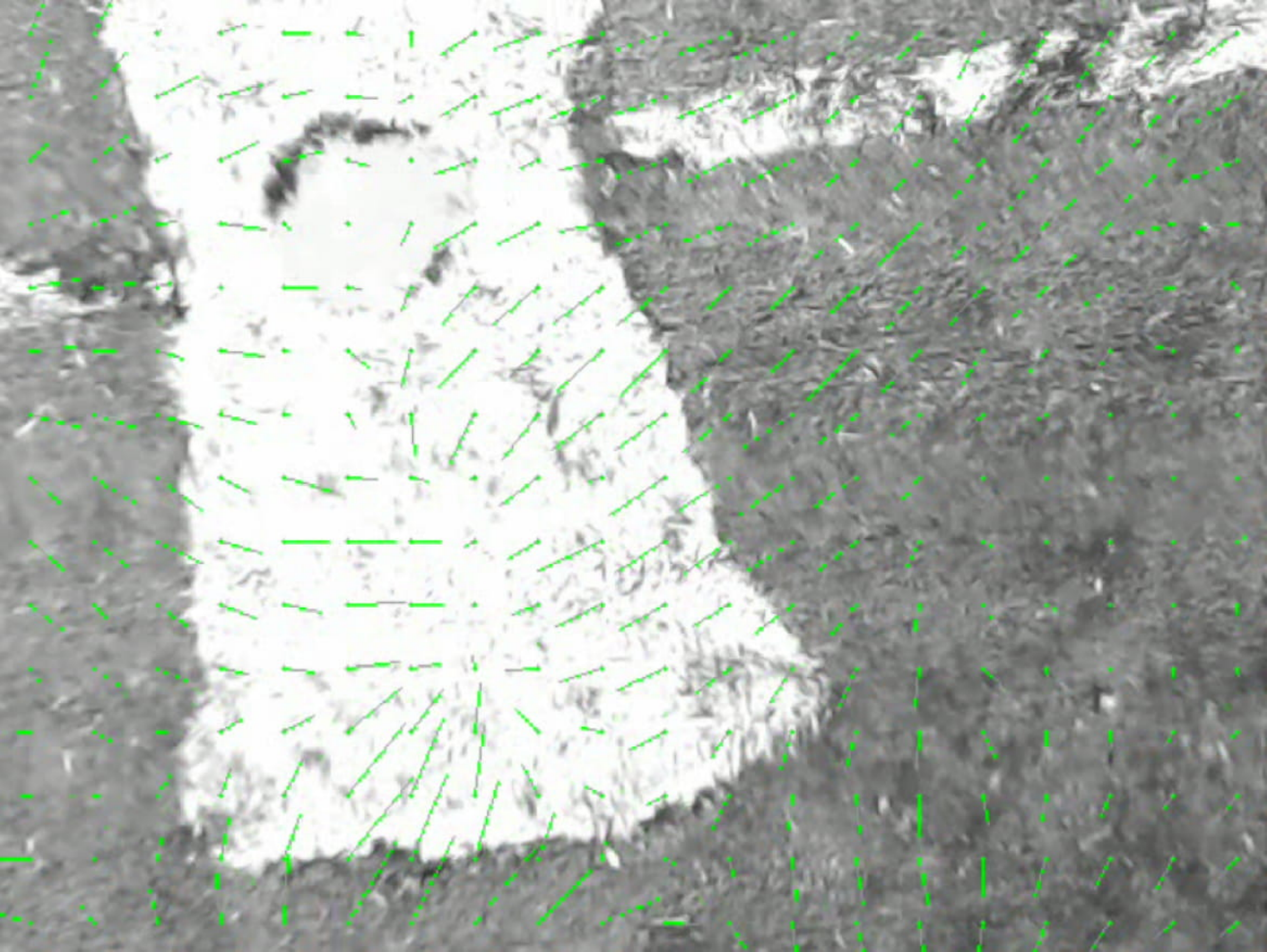}
        \caption{}
	\end{subfigure}
\hfill
	\begin{subfigure}{0.23\textwidth}
    	\includegraphics[width=0.95\textwidth]{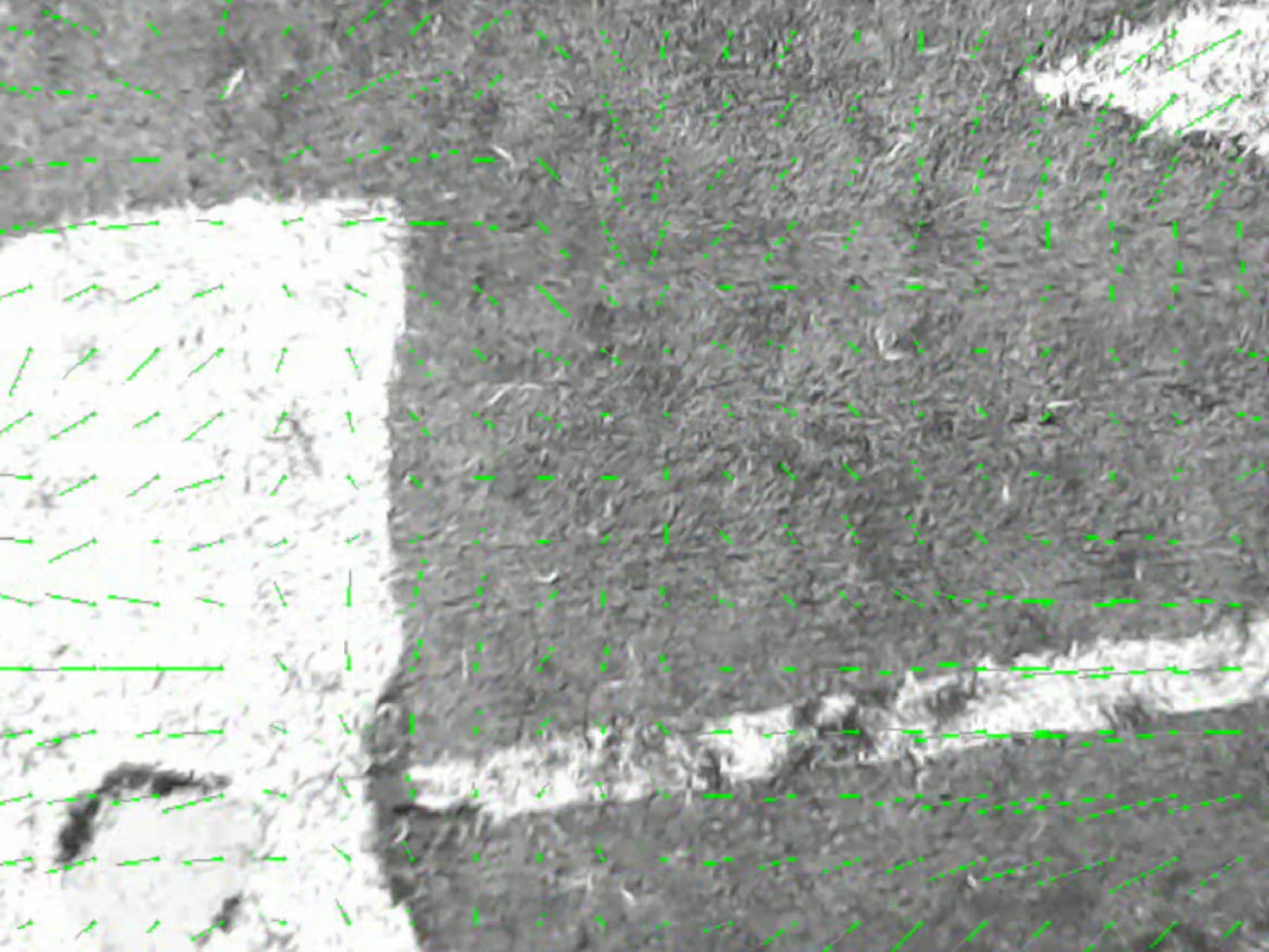}
        \caption{}
	\end{subfigure}
\caption{A sequence of ordered optical flow images taken from the drone. It can be observed in green the estimated velocity vectors.}
\label{fig:of-experiments}
\end{figure}

The Optical Flow (OF) is the pattern of apparent motion of an object between two consecutive frames. This motion can be caused by the movement of the object or the camera trough time. Mathematically, the optical flow consists in a 2D vector field. This 2D motion results from the projection of moving 3D objects in the image plane. Each vector contains the data that shows the movement of the detected objects from one frame to the next, as can be seen at Fig. \ref{fig:of-experiments}. Optical flow method is based in two main assumptions given next:
\begin{thm}
\label{ass:1}
The pixel intensity of a detected object do not change between consecutive frames.
\end{thm}
\begin{thm}
\label{ass:2}
Pixels near to the detected object have similar motions.
\end{thm}
The 2D motion equations establishes that $I(x,y,t)$ is the center voxel (a pixel in three dimension) on an $m \times n$ neighborhood, then it moves by $\Delta x$ and $\Delta y$ in a determined time $\Delta t$ to $(x+\Delta x,y+\Delta y,t+\Delta t)$, so the next equation is assumed \cite{8282853}
\begin{equation}
\label{eq:flow}
I(x,y,t) = I(x+\Delta x,y+\Delta y,t+\Delta t).
\end{equation}
Since the displacements are differential, it implies that they are very small, so the Taylor approach is used around $I(x,y,t)$ to simplify (\ref{eq:flow}) as follows
\begin{eqnarray}
\label{eq:flow-hot}
I(x+\Delta x,y +\Delta y,t+\Delta t) &=& I(x,y,t)+\\ \nonumber
& &\frac{\Delta I}{\Delta x}\Delta x + \frac{\Delta I}{\Delta y}\Delta y +\\ 
& &\frac{\Delta I}{\Delta t}\Delta t + H.O.T. \nonumber
\end{eqnarray}
where $H.O.T.$ means higher-order terms. From (\ref{eq:flow-hot}) and considering Assumptions \ref{ass:1} and \ref{ass:2} we obtain
\begin{equation}
\label{of-equation}
	\frac{\Delta I}{\Delta x}v_x + \frac{\Delta I}{\Delta y}v_y + \frac{\Delta I}{\Delta t} = 0
\end{equation}
where $v_x=\frac{\Delta x}{\Delta t}$ and $v_y=\frac{\Delta y}{\Delta t}$ are the optical flow, or in other words, the $(v_x,v_y)$ are the components of the image velocity, and the $\frac{\Delta I}{\Delta x}, \frac{\Delta I}{\Delta y}, \frac{\Delta I}{\Delta t}$ are the image intensity derivatives of $(x,y,t)$ aka $I_x$, $I_y$ and $I_z$ \cite{OFp}. Clearly $(v_x,v_y)$ are unknown and we cannot solve (\ref{of-equation}) with two unknown variables, for that we use the so-called \textit{Gunnar-Farneback} algorithm.

The Gunnar-Farneback algorithm \cite{farneback} is designed to produce dense optical flow techniques. This technique is based on producing dense grid of points, which is very useful for applications like learning or recognition. It consists in two-frame motion estimation algorithm. This method approximate each neighborhood of both frames by quadratic polynomials. From observing how an exact polynomial transforms under translation, a method to estimate displacement fields from the polynomial expansion coefficients is derived to lead an algorithm. In this paper we omit the details of Gunnar-Farneback algorithm derivation, the reader who is interested on that, please see reference \cite{farneback}.

%
%
%
%
Some experiments were conducted to show the effectiveness of this dense optical flow algorithm. A sequence of four pictures are depicted in Fig. \ref{fig:of-experiments} showing the vectors indicating the OF direction. The complete video of this experiment can be seen at
\newline
\url{https://youtu.be/_0NjOPtnJsU}
\newline
%
\subsubsection{Extended Kalman Filter algorithm}
The Kalman Filter is the most famous method for prediction in systems with random noise. The key idea behind Kalman Filter is presented in two steps: \textit{Prediction} and \textit{Update}. \textit{Prediction step} consists in the prediction of the state variables of the system based on the inputs and previous values of system states. In the \textit{update step}, the prediction is improved using the data of external measurements \cite{kalmanmatlab2}. The Extended Kalman Filter (EKF) is the nonlinear version of the Kalman Filter, since most realistic problems in robotics involves nonlinear functions, in this paper we have implemented EKF for position estimation; we have taken into account the data obtained from the optical flow algorithm to improve UAV position accuracy. In order to implement the EKF we need a mathematical model that represents the quadrotor nonlinear dynamics. First we present the classical Kalman Filter linear model
\begin{equation}
\label{eq:Kalman}
\begin{aligned}
&x_k = A_k x_{k-1} + B_k u_k + w_k \\
&y_k = C_k x_k + v_k
\end{aligned}
\end{equation}
where $x_k$ is the state vector given at time \textit{k}; $y_k$ is the output vector given at time \textit{k}; $u_k$ is the control input at time \textit{k}; $w_k$ and $v_k$ are process noise and measurement noise, respectively; $A_k$ is the transition matrix; $B_k$ is the input matrix; and $C_k$ is the output matrix that describes the mapping from state $x_k$ to an output $y_k$. The nonlinear representation of the Kalman Filter aka EKF is
\begin{equation}
\label{eq:EKF}
\begin{aligned}
&x_k = f(x_{k-1} , u_k) + w_k \\
&y_k = g(x_{k}) + v_k
\end{aligned}
\end{equation}
where \textit{f} and \textit{g} are nonlinear states and output functions respectively. EKF is conformed by subsequent linearization of model (\ref{eq:EKF}) around system's equilibrium point, resulting in a sequence of models similar to (\ref{eq:Kalman}).
For the implementation, the state vector is defined as follows
\begin{itemize}
  \item Quaternions: $(q_0, q_1, q_2, q_3)$.
  \item Velocity (North, East, Down).
  \item Position (North, East, Down).
  \item Delta Angle bias $(x,y,z)$.
  \item Delta Velocity bias.
  \item Wind Vector (North, East).
  \item Earth Magnetic Field Vector (North, East, Down).
  \item Body Magnetic Field Vector $(x,y,z)$.
\end{itemize}
And the output vector consists in:
\begin{itemize}
\item Roll angle $\phi$.
\item Pitch angle $\theta$.
\item Yaw angle $\psi$.
\item Velocities in North, Down and East, $(V_N,V_D,V_E)$.
\item Positions IN North, Down and East relative to UAV takeoff position $(P_N,P_D,P_E)$.
\item $(x,y,z$) gyro biases, $(G_x,G_y,G_z)$.
\end{itemize}
%
\section{Experiments} \label{sec:exp}
In this Section, the experiments are presented. Such experiments are in two steps:
\begin{enumerate}
\item \textit{Hovering flight}. The first part of the experiment consists of making flight tests in hovering mode.
\item \textit{Trajectory tracking flight}. In this experiment the UAV must follow a predefined path given as waypoints.
\end{enumerate}
In the next subsections we explain the obtained results of both aforementioned cases.
%
\subsection{Hovering}
\begin{figure}[h!]
  \centering
\includegraphics[width=0.4\textwidth]{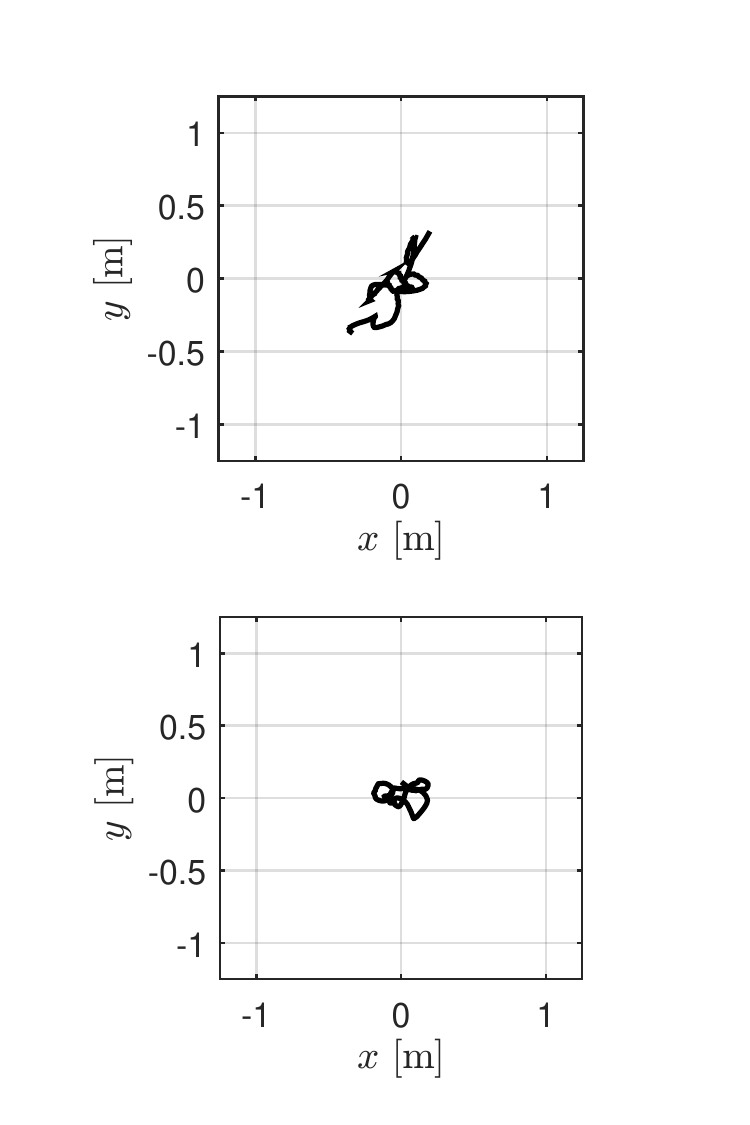}
  \caption {Loiter stability comparison. \textbf{Above}: Loiter mode \textbf{without using DOF algorithm}, only GPS and INS information is fused in the Kalman Filter. \textbf{Below}: Loiter mode \textbf{using DOF algorithm} and GPS-INS information. Both of them have the same axis limits ($1.25$m) to appreciate the differences.}
  \label{loiter}
\end{figure}

The first experiment consists of flying the UAV in the flight mode called hovering, aka \textit{loiter}. When this flight mode is activated, the UAV should keep flying in the actual position in which it is initially located. For comparison purposes we first tested this mode in the quadrotor platform shown at Fig. \ref{drone}, considering only GPS and INS data fusion by using the well-known \textit{Arducopter} firmware, which uses a Kalman Filter to estimate the quadrotor states. A video of the obtained results can be seen at 
\newline
\url{https://youtu.be/FTjoCMh64NQ} 
\newline
The video consists in two clips with the UAV flying in hover mode. The first clip does not use the OF implementation, while the second clip does use it.
Fig. \ref{loiter} shows the performance of the flying test. In the first experiment, the flight was performed without the use of the DOF algorithm, in order to visualize the normal behavior of this flight mode. Regarding the second experiment, it is performed with the optical flow implementation. It can be appreciate smoother movements without considerable variations, so the position error is minimized due the optical flow measurements.

%
\subsection{Trajectory tracking}
\begin{figure}\vspace{0.5cm}
  \centering
  \includegraphics[width=0.4\textwidth]{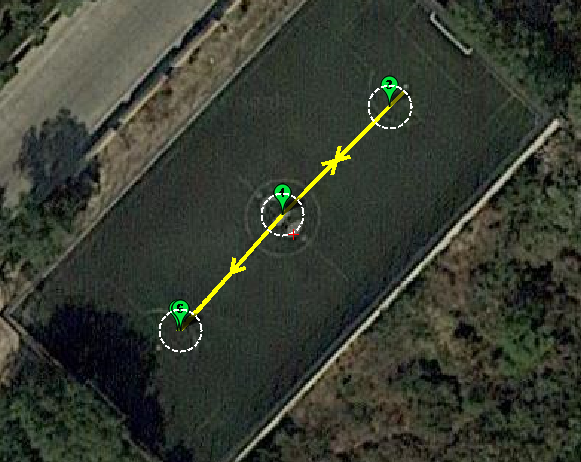}
  \caption{Google Earth view of the desired path. Such path consists of four waypoints located in a soccer field. The quadrotor UAV must take-off automatically from a given waypoint (home position). The first waypoint is in the center of the field and the UAV must go to that waypoint after takeoff. Second waypoint is in the bottom of the image. After reach this waypoint, the UAV turns $180$ degrees and return again to the home position for landing.}
  \label{fig:plan}
\end{figure}

\begin{figure}[h!]
  \centering
    \includegraphics[width=0.48\textwidth]{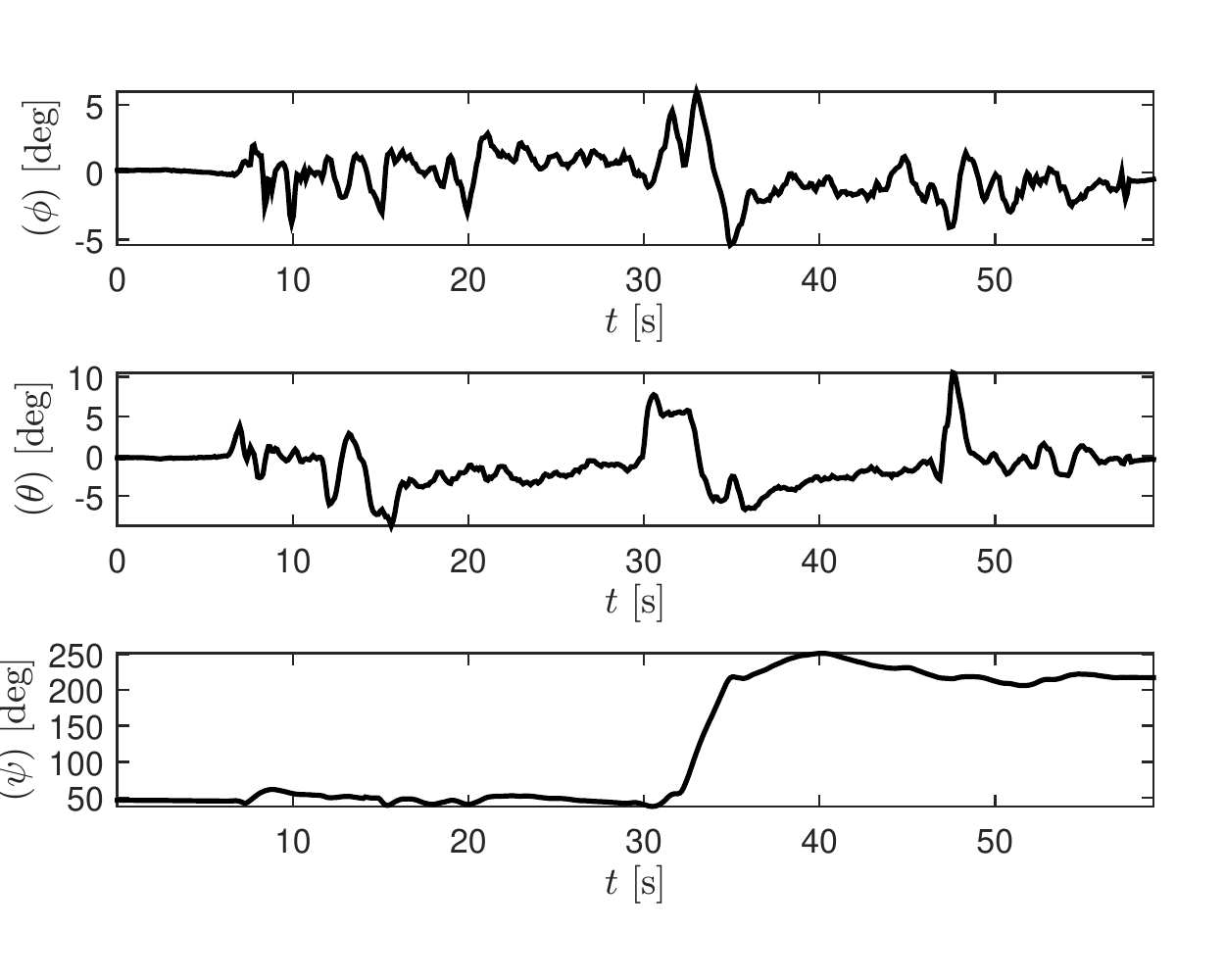}
  \caption{Angle displacement in roll ($\phi$), pitch ($\theta$) and yaw ($\psi$) during the trajectory tracking \textbf{without OF implementation}.}
  \label{rpynoof}
\end{figure}

\begin{figure}[h!]
  \centering
    \includegraphics[width=0.475\textwidth]{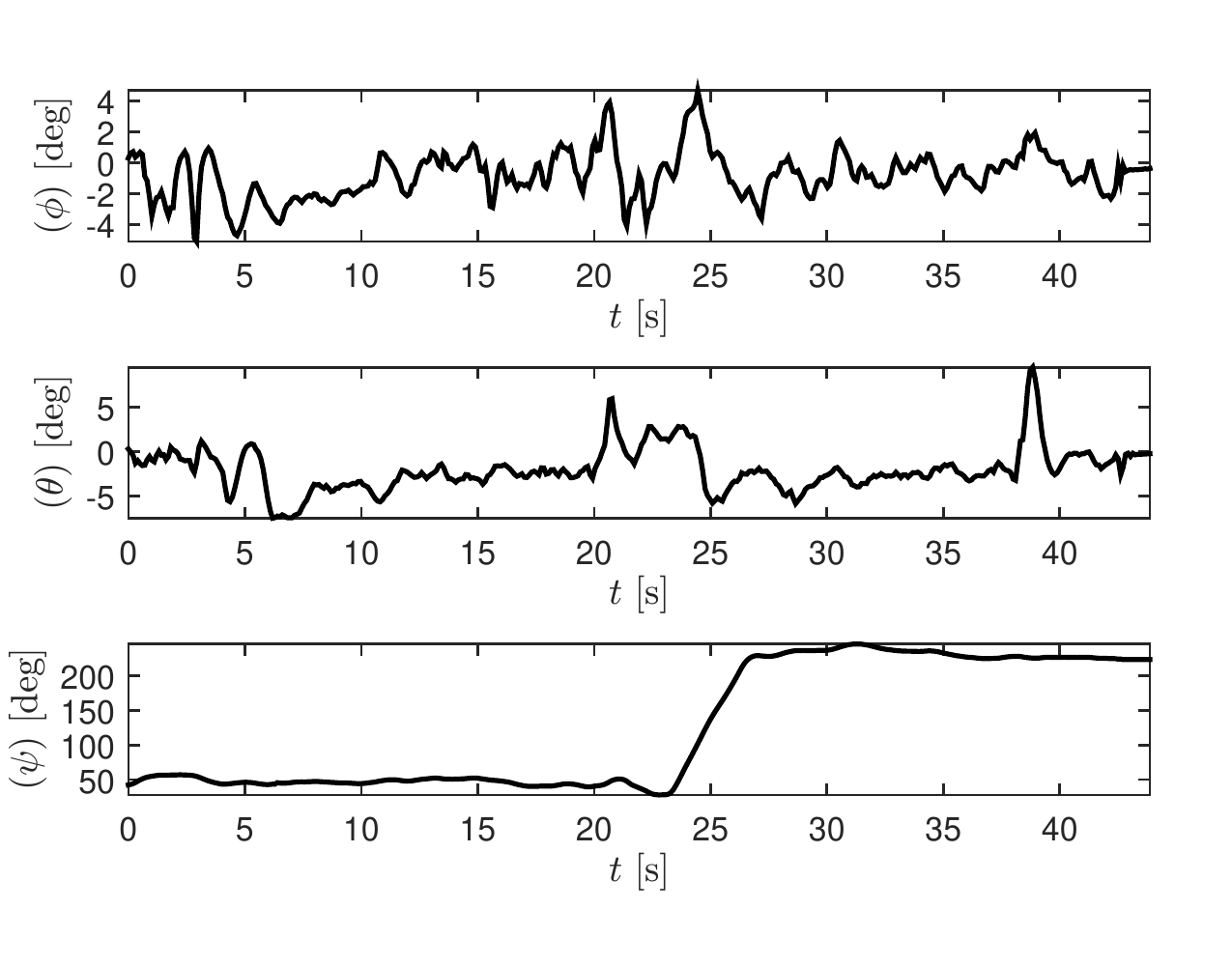}
  \caption{Angle displacement in roll ($\phi$), pitch ($\theta$) and yaw ($\psi$) during the trajectory tracking \textbf{with the OF implementation}.}
  \label{rpyof}
\end{figure}

Now, a given trajectory must be followed by the drone, in order to investigate the performance of the fusion of DOF algorithm together with GPS and INS devices. The trajectory is conformed by a line. Two waypoints are chosen to form the line; also two waypoints indicate take-off and landing position as can be seen at Fig. \ref{fig:plan}. 
To replicate the common way to execute this flight path, first, such flight path was performed by the quadrotor without using dense optical flow algorithm. Some variations in height occurred during the flight path execution, this is due that only GPS and barometer information are fused to estimate drone's altitude. The flight path without optical flow was expected to have en error around $2.5$ meters, this value is very close to the edge of the GPS error indicated by the manufacturer. The roll ($\phi$) and pitch ($\theta$) were in constant variations, showing inability to maintain a fixed course.
Then, the same trajectory was performed but this time with the influence of optical flow data into the EKF. In this test it can be observed that the movements are smoother w.r.t. the common approach. Also, the UAV position is smooth with minimal variations, demonstrating an improvement in the trajectory tracking. This performances can be seen at Figs. \ref{rpynoof} and \ref{rpyof}, respectively. Further, a pair of videos showing the execution of the path can be seen at:
\newline
\url{https://youtu.be/jiyyPWVV3nE} (\textbf{Experiment with no OF implementation}.)
\newline
\url{https://youtu.be/rzISpDJs4t8} (\textbf{Experiment with OF implementation}.)
\newline

\begin{figure}[h!]
  \centering
    \includegraphics[width=0.5\textwidth]{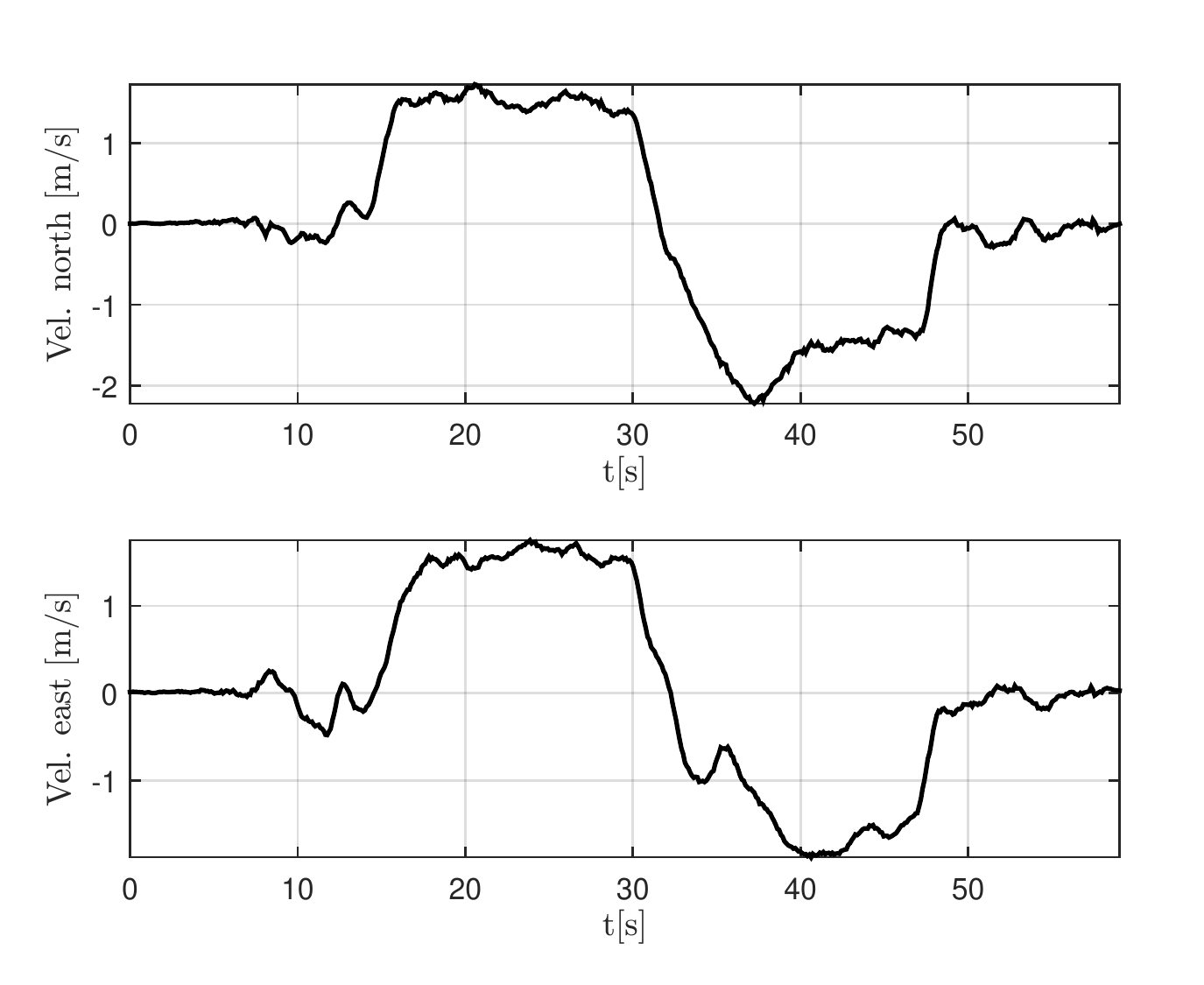}
  \caption{Velocities in the UAV north and east direction during trajectory tracking \textbf{without using OF}.}
  \label{vel-gps}
\end{figure}
\begin{figure}[h!]
  \centering
    \includegraphics[width=0.5\textwidth]{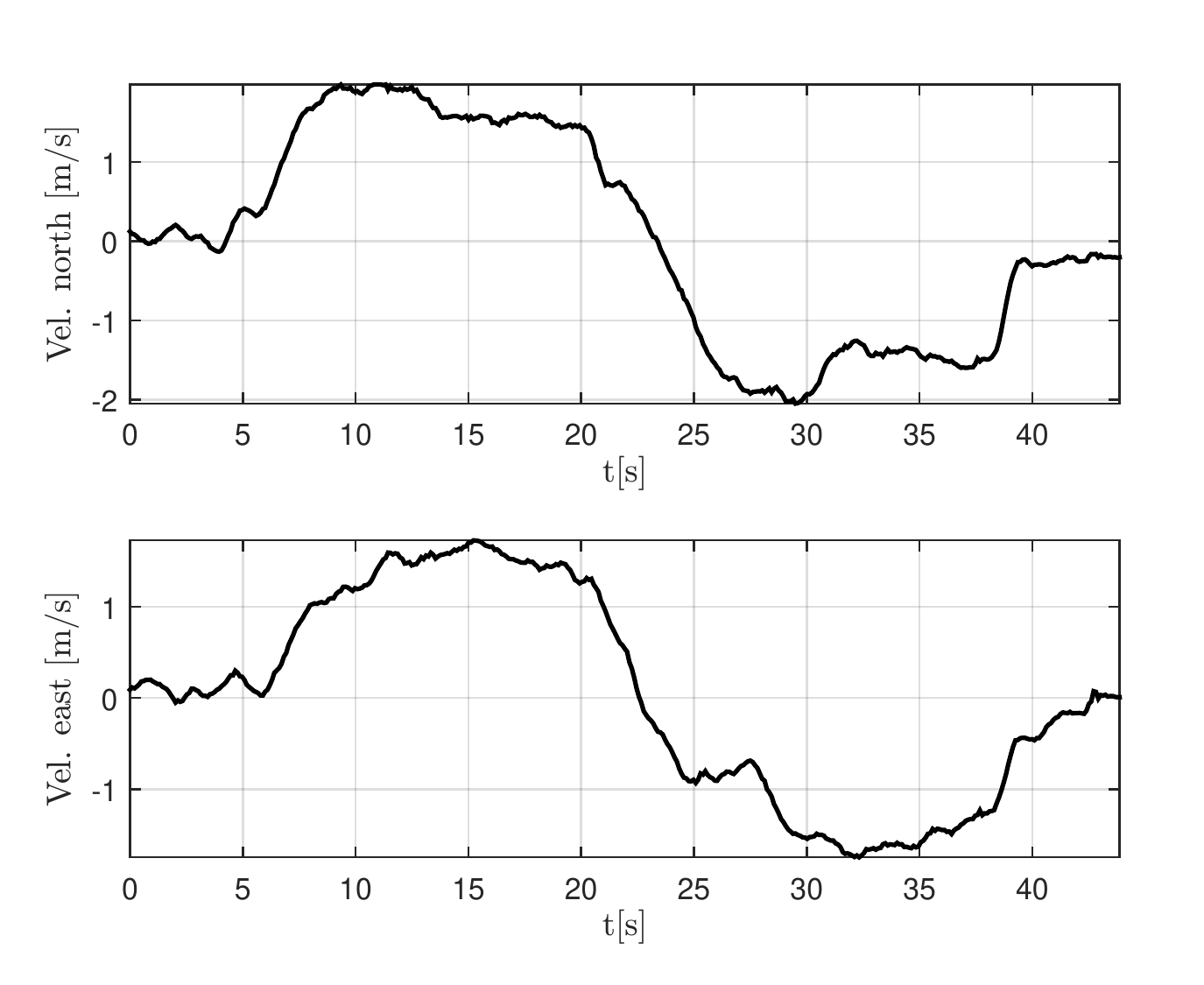}
  \caption{Velocities in the UAV north and east direction during the during the trajectory tracking \textbf{using optical flow}.}
  \label{vel-of}
\end{figure}
\begin{figure*}[htb!]
  \centering
\includegraphics[width=0.95\textwidth,height=7.5cm]{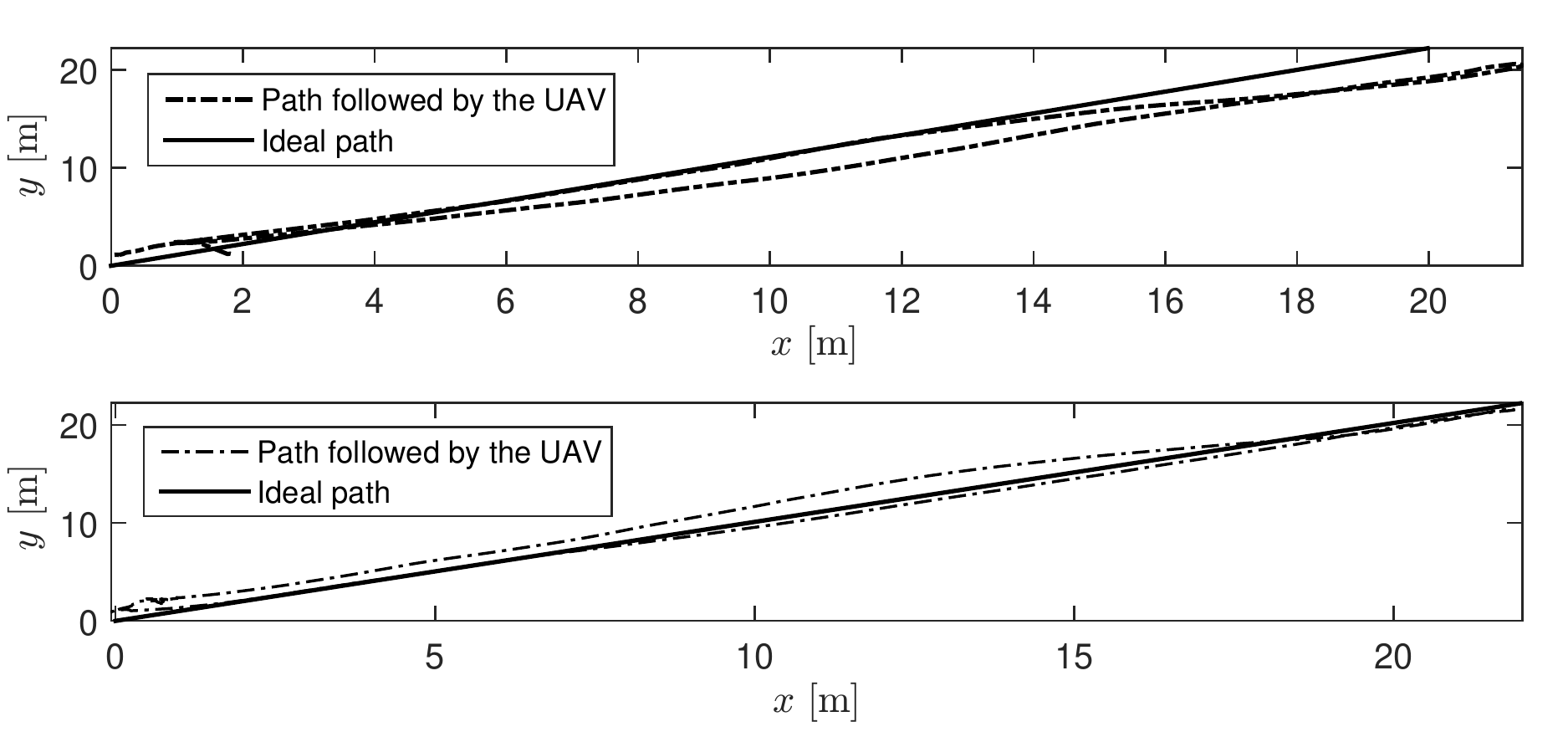}
  \caption{Graphs in the $x-y$ plane showing the UAV position performance during the trajectory tracking. Above, only GPS-INS information used in EKF; below, EKF with optical flow implementation.}
  \label{2d}
\end{figure*}

More interesting are the results in position estimation using dense optical flow algorithm. A graph in 2D is depicted at Fig. \ref{2d} which shows the performance of the proposed approach. In the performance without OF algorithm it can be appreciate a couple of details:
\begin{itemize}
\item The drone take the turn of $180$ degrees beyond the reference trajectory with an approximate difference of $1$m in the $x$ axis. 
\item In the second part of the trajectory, there is a maximum error (difference between reference and actual trajectory) of approximate $2.5$m in the $y$ axis.
\end{itemize}
In the performance with OF algorithm it can be observed that
\begin{itemize}
\item The UAV position in the turn of $180$ degrees is in the same position as the reference trajectory.
\item The maximum error noted is in the $y$ axis during the first part of the path, that error is approximately equal to $1.5$m.
\end{itemize} 

In Figs. \ref{vel-gps} and \ref{vel-of}, the velocities in the north and east directions are shown. Such figures correspond to the cases with and without the use of the OF algorithm. Since the UAV desired trajectory is given as a complete straight line from home position to a predefined point, and then return to the home position, it is expected that in Fig. \ref{vel-of} can be observe a sinusoidal behavior. Fig. \ref{vel-gps} also tries to emulate this behavior, but not as good as Fig. \ref{vel-of} where OF is implemented.

\begin{figure*}[htb!]
\centering
  \includegraphics[width=16cm,height=6.3cm]{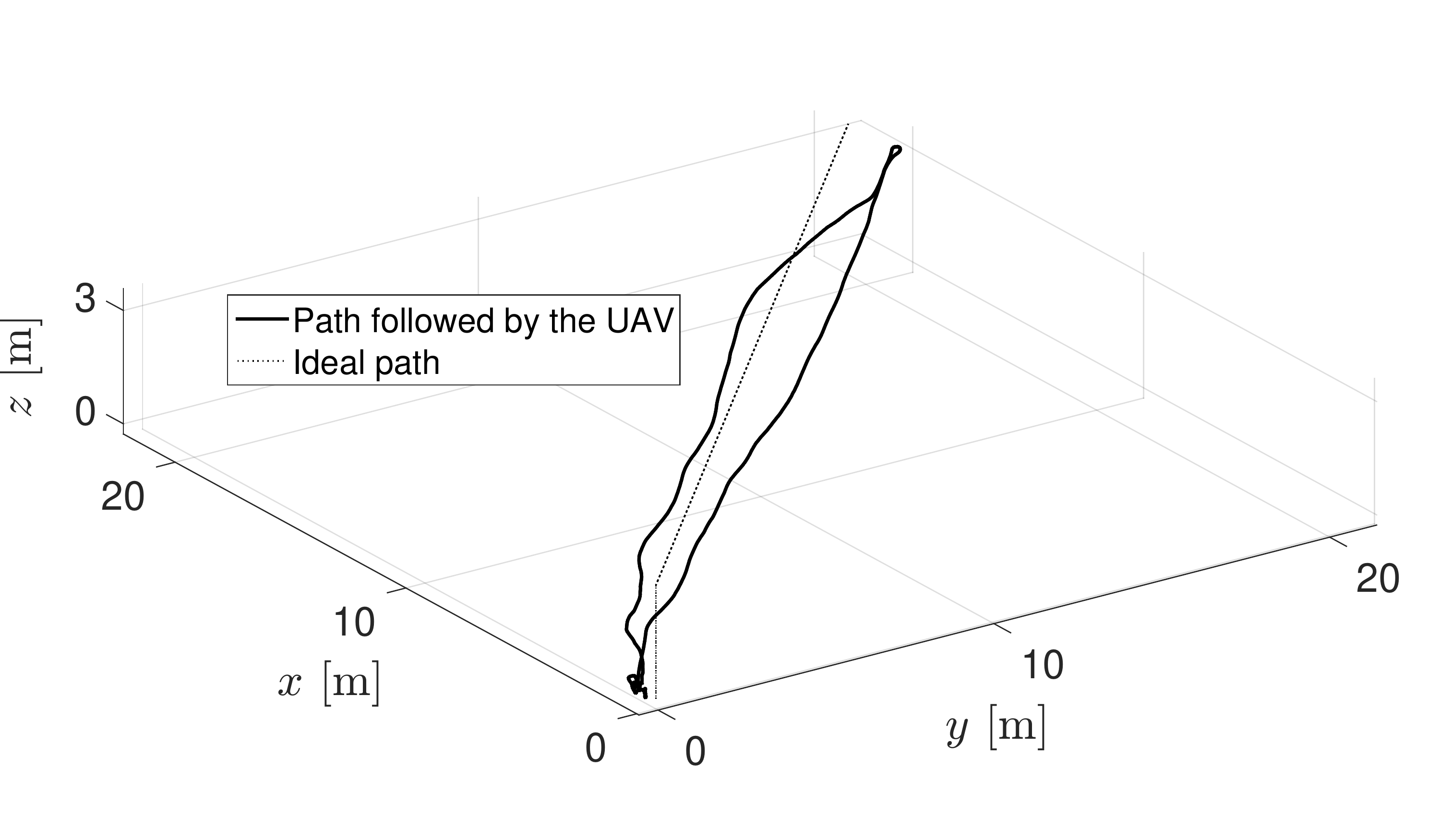}
    \caption{Three-dimensional view of the desired path and actual path performed by the UAV during the trajectory tracking \textbf{without OF algorithm}.}
    \label{pat1}
\end{figure*}

\begin{figure*}[htb!]
\centering
 \includegraphics[width=16cm,height=5.7cm]{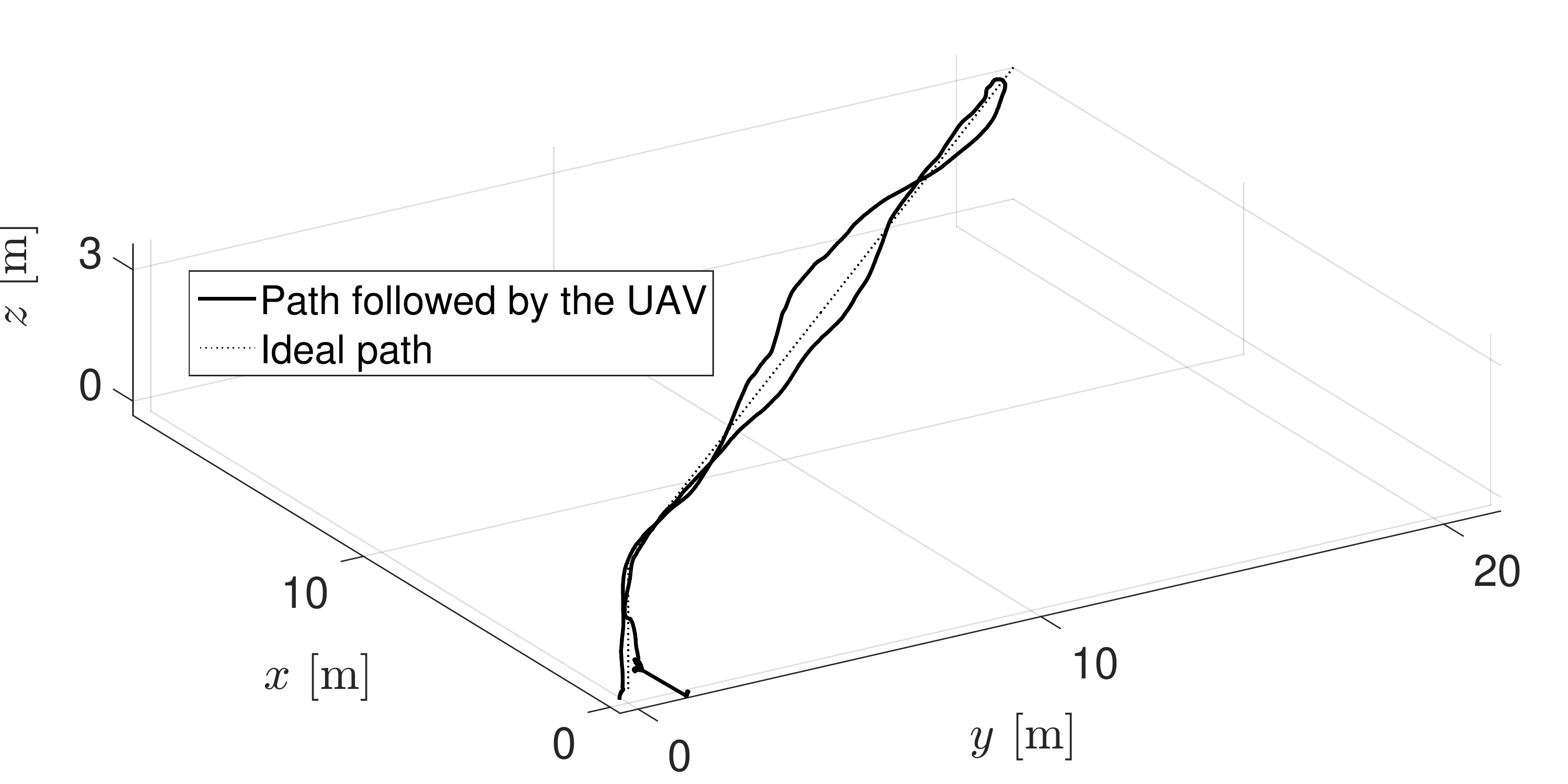}
    \caption{Three-dimensional view of the desired path and actual path performed by the UAV during the trajectory tracking \textbf{with OF algorithm}.}
\label{pat2}
\end{figure*}

Figs. \ref{pat1} and \ref{pat2} shows the three-dimensional view of the paths performed by the UAV. In both cases, using and not using OF algorithm, there was some height variations, this is due to the fact that only information of GPS and barometer is used to estimate quadrotor altitude. As one can see from aforementioned figures, the improvement in the execution of the route when the optical flow is implemented is considerable smoother and more precise than the classical approach.
\section{Conclusions}
\label{sec:conclusion}
The implementation of dense optical flow algorithm has significantly improved the UAV position estimation. Experiments we have conducted demonstrate its effectiveness. Since the quadrotor UAV is a underactuated system, the movements in roll and pitch were smoother when the optical flow is implemented. The yaw movements are very similar in both approaches, as one can see at Figs. \ref{rpynoof} and \ref{rpyof}, since it is only used to initiate the return trajectory. Two flight modes have been tested: hovering aka loiter and trajectory tracking. In both cases the drone presented minimal error with the use of the OF algorithm compared to the conventional EKF approaches where no optical flow algorithm is implemented. As a consequence the quadrotor flies more stable in $x-y$ position.

Future works include: a) improving the UAV altitude estimation adding lidar or vision-based altitude systems; b) testing the proposed approach in fixed-wing aircraft; and c) compare the results in an specific application, for instance: orthomosaics generations and NDVI indexes for precision agriculture.
\section*{Acknowledgments}
The authors acknowledge to National Council of Science and Technology in Mexico (CONACYT) for supporting this research through the FORDECYT-CONACYT program under agreement 000000000292399.\\

\bibliographystyle{IEEEtran}
\bibliography{bibliography}

\begin{thebibliography}{10}
\providecommand{\url}[1]{#1}
\csname url@rmstyle\endcsname
\providecommand{\newblock}{\relax}
\providecommand{\bibinfo}[2]{#2}
\providecommand\BIBentrySTDinterwordspacing{\spaceskip=0pt\relax}
\providecommand\BIBentryALTinterwordstretchfactor{4}
\providecommand\BIBentryALTinterwordspacing{\spaceskip=\fontdimen2\font plus
\BIBentryALTinterwordstretchfactor\fontdimen3\font minus
  \fontdimen4\font\relax}
\providecommand\BIBforeignlanguage[2]{{%
\expandafter\ifx\csname l@#1\endcsname\relax
\typeout{** WARNING: IEEEtran.bst: No hyphenation pattern has been}%
\typeout{** loaded for the language `#1'. Using the pattern for}%
\typeout{** the default language instead.}%
\else
\language=\csname l@#1\endcsname
\fi
#2}}

\bibitem{7500933}
P.~Henkel and A.~Sperl, ``Real-time kinematic positioning for unmanned air
  vehicles,'' in \emph{2016 IEEE Aerospace Conference}, March 2016, pp. 1--7.

\bibitem{8115945}
S.~Takahashi, N.~Kubo, N.~Yamaguchi, and T.~Yokoshima, ``Real-time monitoring
  for structure deformations using hand-held rtk-gnss receivers on the wall,''
  in \emph{2017 International Conference on Indoor Positioning and Indoor
  Navigation (IPIN)}, Sept 2017, pp. 1--7.

\bibitem{4543471}
R.~He, S.~Prentice, and N.~Roy, ``Planning in information space for a quadrotor
  helicopter in a gps-denied environment,'' in \emph{2008 IEEE International
  Conference on Robotics and Automation}, May 2008, pp. 1814--1820.

\bibitem{5152680}
S.~Ahrens, D.~Levine, G.~Andrews, and J.~P. How, ``Vision-based guidance and
  control of a hovering vehicle in unknown, gps-denied environments,'' in
  \emph{2009 IEEE International Conference on Robotics and Automation}, May
  2009, pp. 2643--2648.

\bibitem{7738795}
S.~E. Tsai and S.~H. Zhuang, ``Optical flow sensor integrated navigation system
  for quadrotor in gps-denied environment,'' in \emph{2016 International
  Conference on Robotics and Automation Engineering (ICRAE)}, Aug 2016, pp.
  87--91.

\bibitem{WCICA14:Hong}
Y.~Hong, X.~Lin, Y.~Zhuang, and Y.~Zhao, ``Real-time pose estimation and motion
  control for a quadrotor uav,'' in \emph{Proceeding of the 11th World Congress
  on Intelligent Control and Automation}, June 2014, pp. 2370--2375.

\bibitem{ACC13:Chao}
H.~Chao, Y.~Gu, J.~Gross, G.~Guo, M.~L. Fravolini, and M.~R. Napolitano, ``A
  comparative study of optical flow and traditional sensors in uav
  navigation,'' in \emph{2013 American Control Conference}, June 2013, pp.
  3858--3863.

\bibitem{IROS14:Rhudy}
M.~B. Rhudy, H.~Chao, and Y.~Gu, ``Wide-field optical flow aided inertial
  navigation for unmanned aerial vehicles,'' in \emph{2014 IEEE/RSJ
  International Conference on Intelligent Robots and Systems}, Sept 2014, pp.
  674--679.

\bibitem{Fusini:ICUAS15}
L.~Fusini, J.~Hosen, H.~H. Helgesen, T.~A. Johansen, and T.~I. Fossen,
  ``Experimental validation of a uniformly semi-globally exponentially stable
  non-linear observer for gnss- and camera-aided inertial navigation for
  fixed-wing uavs,'' in \emph{2015 International Conference on Unmanned
  Aircraft Systems (ICUAS)}, June 2015, pp. 851--860.

\bibitem{Qi:ICUS17}
J.~Qi, N.~Yu, and X.~Lu, ``A uav positioning strategy based on optical flow
  sensor and inertial navigation,'' in \emph{2017 IEEE International Conference
  on Unmanned Systems (ICUS)}, Oct 2017, pp. 81--87.

\bibitem{ICRA13:Flow}
D.~Honegger, L.~Meier, P.~Tanskanen, and M.~Pollefeys, ``An open source and
  open hardware embedded metric optical flow cmos camera for indoor and outdoor
  applications,'' in \emph{2013 IEEE International Conference on Robotics and
  Automation}, May 2013, pp. 1736--1741.

\bibitem{tractor}
M.~Perez-Ruiz, D.~C. Slaughter, C.~Gliever, and S.~K. Upadhyaya,
  ``Tractor-based real-time kinematic-global positioning system (rtk-gps)
  guidance system for geospatial mapping of row crop transplant,''
  \emph{Biosystems Engineering}, vol. 111, no.~1, pp. 64 -- 71, 2012.

\bibitem{ppkvsrtk}
\BIBentryALTinterwordspacing
``Precise positioning system for uavs.'' [Online]. Available:
  \url{https://www.suasnews.com/2016/05/43820/}
\BIBentrySTDinterwordspacing

\bibitem{Mercado:ICUAS13}
D.~A. Mercado, G.~Flores, P.~Castillo, J.~Escareno, and R.~Lozano,
  ``Gps/ins/optic flow data fusion for position and velocity estimation,'' in
  \emph{2013 International Conference on Unmanned Aircraft Systems (ICUAS)},
  May 2013, pp. 486--491.

\bibitem{stgps}
U-blox, ``Neo-7 u-blox 7 gnss modules data sheet,'' U-blox, USA, Tech. Rep.
  UBX-13003830-R07, Nov. 2014.

\bibitem{8282853}
K.~Noor, E.~A. Siddiquee, D.~Sarma, A.~Nandi, S.~Akhter, S.~Hossain,
  K.~Andersson, and M.~S. Hossain, ``Performance analysis of a surveillance
  system to detect and track vehicles using haar cascaded classifiers and
  optical flow method,'' in \emph{2017 12th IEEE Conference on Industrial
  Electronics and Applications (ICIEA)}, June 2017, pp. 258--263.

\bibitem{OFp}
D.~Patel and S.~Upadhyay, ``Optical flow measurement using lucas kanade
  method,'' \emph{International Journal of Computer Applications}, vol.~61,
  no.~10, pp. 6 -- 10, 2013.

\bibitem{farneback}
G.~Farneb{\"a}ck, ``Two-frame motion estimation based on polynomial
  expansion,'' in \emph{Image Analysis}, J.~Bigun and T.~Gustavsson, Eds.\hskip
  1em plus 0.5em minus 0.4em\relax Berlin, Heidelberg: Springer Berlin
  Heidelberg, 2003, pp. 363--370.

\bibitem{kalmanmatlab2}
R.~G. Brown and P.~Y.~C. Hwang, \emph{Introduction to Random Signals and
  Applied Kalman Filtering with Matlab Exercises}, 4th~ed.\hskip 1em plus 0.5em
  minus 0.4em\relax Wiley-IEEE Press, 2012.

\end{thebibliography}
\end{document}